\documentclass[12pt]{article}
\usepackage{graphicx}

\newsavebox{\sboxpubnumber}
\newsavebox{\sboxpubdate}
\newcommand{\pubdate}[1]{\begin{lrbox}{\sboxpubdate}{#1}\end{lrbox}}
\newcommand{\pubnumber}[1]{\begin{lrbox}{\sboxpubnumber}{\begin{tabular}{l} #1 \\
				 \usebox{\sboxpubdate}
				 \end{tabular}}
                           \end{lrbox}
                           \pubblock}
\newcommand{\Title}[1]{\begin{center} {\Large #1 } \end{center}}
\newcommand{\Author}[1]{\begin{center}{ \sc #1} \end{center}}
\newcommand{\Address}[1]{\begin{center}{ \it #1} \end{center}}

\newcommand{\pubblock}{\rightline{
			\usebox{\sboxpubnumber}}}
\newenvironment{Abstract}{\begin{quotation}  }{\end{quotation}}
\newenvironment{Presented}{\begin{quotation} \begin{center}
             PRESENTED AT\end{center}\bigskip
      \begin{center}\begin{large}}{\end{large}\end{center}
      \end{quotation}}
\newcommand{\Acknowledgements}{\bigskip  \bigskip \begin{center} \begin{large}
             \bf ACKNOWLEDGEMENTS \end{large}\end{center}}
\newcommand{\comm}[1]{}   
\newcommand{\gsim}{\raisebox{-0.8ex}{\mbox{$\stackrel{\textstyle>}{\sim}$}}}

\newcommand{\ncs}{N_{_{\rm CS}}}

\newcommand{\nw}{N_{\rm wind}}

\newcommand{\be}{\begin{equation}}
\newcommand{\ee}{\end{equation}}

\begin{document}

\begin{titlepage}
\pubdate{December 4, 2001}                    
\pubnumber{hep-ph/0112053} 

\vfill
\Title{Topological transitions at the resonance in gauge sector}
\vfill
\Author{Dmitri Grigoriev \footnote{On leave of absence from
Institute for Nuclear Research of Russian Academy of Sciences} }
\Address{Dept.~of Mathematical Physics, National
University of Ireland, \\
Maynooth, Co. Kildare, Ireland}
\vfill
\begin{Abstract}
We discuss topological transitions during parametric resonance in the
gauge sector of electroweak theory. It is shown that the resonance
leads to separation of topological indices of the gauge and
Higgs fields, resulting in topological transitions of non-sphaleron
nature.  
\end{Abstract}
\vfill
\begin{Presented}
    COSMO-01 \\
    Rovaniemi, Finland, \\
    August 29 -- September 4, 2001
\end{Presented}
\vfill
\end{titlepage}
\def\thefootnote{\fnsymbol{footnote}}
\setcounter{footnote}{0}

\section{Introduction}

	It has been suggested recently \cite{Cornwall:2001hq} that
parametric resonance in the gauge sector of electroweak theory can
substantially amplify baryoproduction due to CP-asymmetry present in
the effective action and/or initial conditions. For many electroweak
baryogenesis models,
the baryoproduction is typically the product of CP-asymmetry times the
mean shift of Chern-Simons number $\ncs$ during out of equilibrium stage. In
our scenario, $\ncs$ becomes extremely
high due to the resonance \cite{Cornwall:2000eu,Cornwall:2001hq,alex},
getting parametrically close to unit charge per one sphaleron volume.

To get more specific estimate of baryoproduction
\cite{Cornwall:2001hq}, one needs to understand the dynamics of
topology-changing processes in detail.  In this talk we concentrate on
the behaviour of topological indices during the parametric resonance
in the gauge sector. We show that in our case the mechanism of
topological transitions is considerably more complicated than in other
scenarios of electroweak baryogenesis.  Namely, while topological
transitions changing both the Chern-Simons number of the gauge field
and winding number of the Higgs field take place, during the resonance
they never\footnote{To be precise, normal sphaleron transitions can
occur, but they give zero contribution to the net baryoproduction.}
follow standard sphaleron path found in
\cite{Manton:1983nd,Klinkhamer:1984di}.

\begin{figure}[htb]
    \centering \includegraphics[height=3.5in]{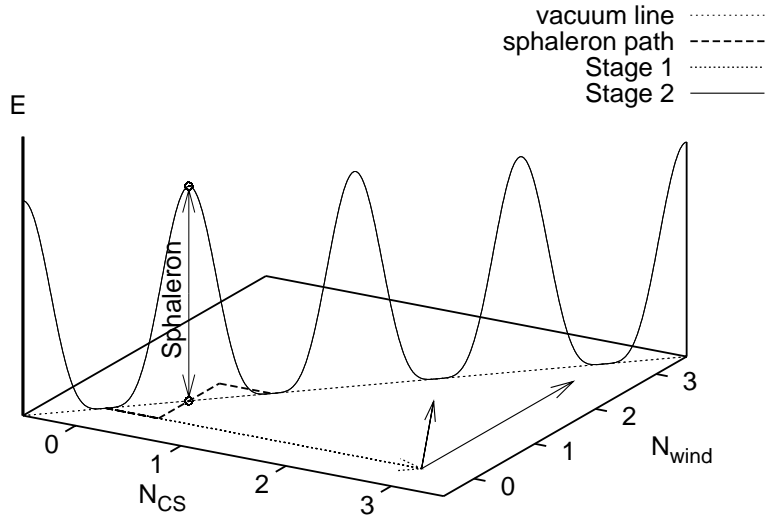}
    \caption{On $(\ncs,\nw)$ plane, the standard periodical vacuum
    structure exists along the vacuum line $\ncs=\nw$; other directions give
    quite different energy profiles -- see Ref.~\cite{Cornwall:2001hq}
    and Figs.~\ref{fig2},\ref{fig3} below. During normal sphaleron
    transition, the system moves to the neighbouring vacuum via the
    sphaleron path, while in our case transition paths are more
    complicated (arrows). }
    \label{fig1}
\end{figure}

This point is illustrated at Fig.~\ref{fig1}. The system states
with minimal energies are located along the vacuum line $\ncs=\nw$;
in most existing electroweak baryogenesis scenarios CP-violating
effects are driving both $\ncs$ and $\nw$ along this line, with
transitions between neighbouring topological vacua going along
normal sphaleron path.

Unlike this common situation, the resonant amplification is a two-step
process. Only the gauge field is directly involved into the resonance,
so at first the system gets through the resonance into highly excited
state with large topological index of gauge field (Chern-Simons number
$\ncs$) and small topological index of the Higgs field (winding number
$\nw$, which is exactly 0 for ansatz used in numerical simulations of
Ref.~\cite{Cornwall:2001hq}). On $(\ncs,\nw)$ plane, Fig.~\ref{fig1},
this state is quite a distance away from the line of vacuum states
$\ncs=\nw$. Of course, a lot of fermions is already created at this
moment via the triangle anomaly due to large shift in $\ncs$; however,
none of them would survive the end of the resonance without
topological transitions in Higgs sector which take place at the second
stage of the process.

The second stage corresponds to the decay of the topological state
created through the resonance. The most favourable path from the point
with large $\ncs$ and zero $\nw$ is obviously back along $\nw=0$ line,
but the gauge field dynamics in this direction is controlled by the
resonance itself, and no transition at constant $\nw$ would create
final state fermions anyway. However, any movement towards the vacuum
line $\ncs=\nw$ is energetically favourable; for relatively
small $\ncs-\nw$, winding number-changing transitions must go over
energy barrier (e.g. through thermal or non-thermal activation), while
at very large $\ncs$ densities typical for gauge-sector resonance
such transitions may become unsuppressed.

\section{Stage I: the resonance in gauge sector}

Homogeneous field ansatz used in
\cite{Cornwall:2000eu,Cornwall:2001hq} makes things very simple here:
gauge field $A_j=M_{\rm W}\phi\tau_j/{2i}$ is parametrised by single
variable $\phi(t)$ with parabolic-like potential \be\label{ephi}
E(\phi)={M_{\rm W} \over {2g^2}}\left(\phi^4+\phi^2\right) \ee
Anharmonic term in (\ref{ephi}) limits the saturation amplitude of the
resonance to be $\phi_{\rm max} \sim 1$, which corresponds to very
high Chern-Simons number density of order of $0.1$ per one sphaleron
volume (provided the inflaton field has sufficient energy
reserve). However, this density will turn into 0 and then change sign
for every period of $\phi$ oscillations during the resonance. The end
of the resonance would result in $\phi$ finally coming back to its
zero vacuum value with zero Chern-Simons density, what makes all
fermionic states created during the resonance to disappear.

\begin{figure}[htb]
    \centering \includegraphics[height=2.5in]{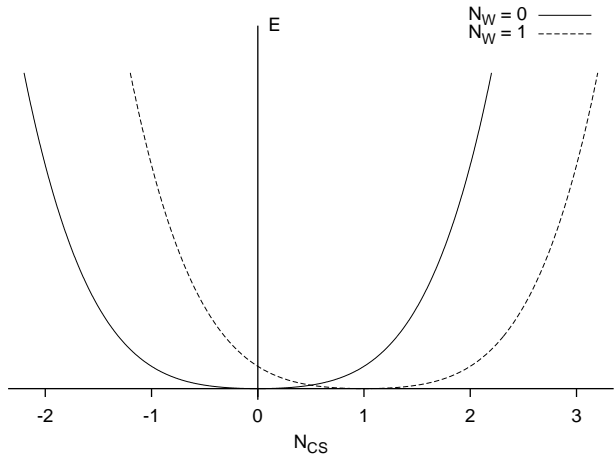}
    \caption{Energy profiles at fixed $\nw$ can be obtained
    \cite{Cornwall:2001hq} directly from Eq.~(\ref{ephi}). To get
    non-zero $\ncs$ after the resonance, one has to change Higgs
    winding number (dashed line).}  \label{fig2}
\end{figure}

Therefore, creation of permanent fermionic states is equivalent to
shifting the minimum point of potential $E(\ncs)$ from 0, see
Fig.~\ref{fig2}. To do that, we need to change winding number of the
Higgs field, once $\ncs=\nw$ at the minimum. As we'll show below,
topological transitions naturally occur in the Higgs sector at
non-zero Chern-Simons number, although their dynamics is hard to
investigate in detail.

\section{Stage II: topological transitions changing the Higgs winding
number}
	
A general understanding of the
dynamics of Higgs winding number-changing topological transitions can
be obtained through the use of gauge invariance. Key argument here is
that the total energy of the bosonic (gauge and Higgs) fields is
invariant under large gauge transformations, which simultaneously
change $\ncs$ and $\nw$ by integer number $N$:
$$
E(\ncs,\nw)=E(\ncs+N,\nw+N)
$$ 
and, therefore,
\be\label{eminus1}
E(\ncs,\nw)=E(\ncs-\nw,0)
\ee 

Note that the function $E(\ncs,0)$ is known through Eq.~(\ref{ephi})
only for homogeneous gauge field, while the large gauge
transformations generally break the homogeneous ansatz. This isn't an
obstacle in (1+1) dimensions, where any field configuration can be
transformed into gauge-equivalent one with homogeneous gauge field;
however, one could expect that the subsequent analysis is also
qualitatively applicable to (3+1) case at non-zero density of
topological numbers, i.e. when the effects due to spatial localization
of field configuration with unit topological charge are negligible.

\begin{figure}[htb]
    \centering
    \includegraphics[height=2.5in]{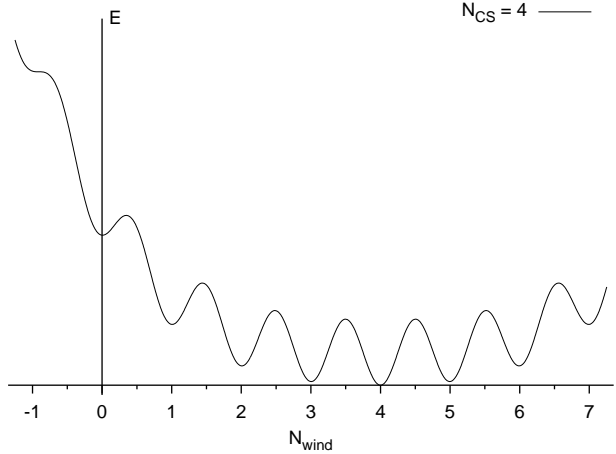}
    \caption{Energy profile at fixed $\ncs$ is in principle identical
    to Fig.~\ref{fig2}, see Eq.~(\ref{eminus1}). However, integer
    values of $\nw$ are separated by sphaleron-like energy barriers;
    we expect them to disappear at large $\left| \ncs-\nw \right|$.}
    \label{fig3}
\end{figure}

Cross-section of $E(\ncs,\nw)$ in the direction orthogonal to
Fig.~\ref{fig2} makes it obvious that non-zero Chern-Simons number
indeed stimulates topological transitions in Higgs sector, once it
is energetically favourable to reduce the difference between $\ncs$
and $\nw$ (see Fig.~\ref{fig3}). Of course, $\nw$ can take only
integer values, so the continuous lines on Fig.~\ref{fig3} just
demonstrate the presence of energy barriers that separate the
neighbouring values of $\nw$. At $\ncs=\nw$ the height of this barrier
should be close to, but not identical to the sphaleron energy, once at
Fig.~\ref{fig3} we are crossing the barrier at fixed integer $\ncs$,
while the sphaleron has half-integer $\ncs$. Although the rigorous
calculation of barrier heights and corresponding transition amplitudes
at non-equal Chern-Simons and winding numbers is yet to be done even
in (1+1) dimensions, it is natural to expect that the barrier in $\nw$
direction should vanish when the energy gain due to transition equals
the sphaleron energy.

This means that topological transitions in Higgs sector should become
very intensive at sufficiently large $\ncs$; one can show
\cite{Cornwall:2001hq} that it happens at $\phi\gsim 1$, i.e.~at
saturation amplitude of parametric resonance.

\section{Conclusions}

The main outcome of semiqualitative arguments presented above is that
the parametric resonance in gauge sector moves us away from well-known
paths in topological space. At both stages of the process the system
never moves along the normal sphaleron path. At the first stage, the
resonance drives the Chern-Simons number along the path with fixed
winding number. Then the increase in $\ncs$ makes the transitions with
$\Delta\nw$ of the same sign more and more favourable until the
transition towards the vacuum line finally takes place. The same
two-stage process will happen at the next half-period of $\ncs$
oscillations, when both $\ncs$ and $\Delta\nw$ will have negative
values. Even for one half-period, the actual trajectory of the system
in topological space cannot be determined from static energy
profiles. Improving estimates given in \cite{Cornwall:2001hq} for
baryoproduction at parametric resonance in gauge sector will thus
inevitably require better understanding of topological transitions at
non-equal Chern-Simons and winding numbers, and more detailed
numerical studies of their kinetics as well. Work in this direction is
now in progress.

\Acknowledgements

This work has been done in collaboration with J.~M.~Cornwall and
A.~Kusenko. The author is also grateful to J.~Garc\'ia-Bellido,
S.~Habib, K.~Heitmann, A.~Rajantie, M.~Shaposhnikov and J.~Smit for
stimulating discussions during COSMO-01, and Organizing Committee of
the conference for support.


\begin{thebibliography}{99}


\bibitem{Cornwall:2000eu}
J.~M.~Cornwall and A.~Kusenko,
Phys.\ Rev.\ D {\bf 61} (2000) 103510
[arXiv:hep-ph/0001058].

\bibitem{Cornwall:2001hq}
J.~M.~Cornwall, D.~Grigoriev and A.~Kusenko,
Phys.\ Rev.\ D {\bf 64} (2001) 123518
[arXiv:hep-ph/0106127].

\bibitem{alex}
A.~Kusenko, hep-ph/0112009 (these Proceedings).

\bibitem{Manton:1983nd}
N.~S.~Manton,
Phys.\ Rev.\ D {\bf 28} (1983) 2019.

\bibitem{Klinkhamer:1984di}
F.~R.~Klinkhamer and N.~S.~Manton,
Phys.\ Rev.\ D {\bf 30} (1984) 2212.

\end{thebibliography}
\end{document}